\documentclass{ws-ppl}
\usepackage{graphicx}
\usepackage[ruled]{algorithm}
\usepackage{algpseudocode}
\usepackage{color}
\begin{document}

\markboth{Parallel Processing Letters}
{Using CFinder on a grid}

%
\catchline{}{}{}{}{}
%

\title{PARALLEL CLUSTERING WITH CFINDER}


%
%

\author{PETER POLLNER}

\address{Statistical and Biological Physics Research Group,
 Hungarian Academy of Sciences, 
Pazmany Peter s 1/a, Budapest, 1117, Hungary,
pollner@hal.elte.hu\\[10pt]
}

\author{GERGELY PALLA}

\address{Statistical and Biological Physics Research Group,
 Hungarian Academy of Sciences, 
Pazmany Peter s 1/a, Budapest, 1117, Hungary,\\[10pt]
}

\author{TAMAS VICSEK}

\address{Department of Biological Physics, Eotvos University,
Pazmany Peter s 1/a, Budapest, 1117, Hungary,\\[10pt]
}

\maketitle

\begin{history}
\received{September 2011}
\revised{November 2011}
\comby{Guest Editors}
 Electronic version of an article published as \\
 PARALLEL PROCESSING LETTERS (PPL) 22:(1) p. 1240001. (2012)\\
 http://dx.doi.org/10.1142/S0129626412400014\\
 copyright World Scientific Publishing Company\\ 
 http://www.worldscinet.com/ppl/22/2201/S0129626412400014.html
\end{history}

\begin{abstract}
The amount of available data about complex systems is increasing every
year, measurements of larger and larger systems are collected and
recorded. A natural representation of such data is given by networks,
 whose size is following the size of the original system.
The current trend of multiple cores in computing
infrastructures call for a parallel reimplementation of earlier methods.
Here we present the grid version of CFinder,
which can locate overlapping communities in directed, weighted or
undirected networks based on the clique percolation method (CPM).
We show that the computation of the communities
can be distributed among several CPU-s or computers. Although switching
to the parallel version not necessarily leads to gain in computing 
time, it definitely makes 
the community structure of extremely large networks accessible.
\end{abstract}


\section{Introduction}
\label{intro}

Many complex systems in nature and society can be successfully
represented in terms of networks capturing the intricate web of
connections among the units they are made of \cite{networkreview}. 
In recent years, several large-scale properties of real-world webs have
been uncovered, e.g., a low average distance combined with a high
average clustering coefficient \cite{Watts-Strogatz}, the broad, scale-free
distribution of node degree \cite{Laci_science,Boccaletti}
and various signatures of hierarchical and modular 
organization \cite{Ravasz02}. 

Beside the mentioned global characteristics, there has been a 
quickly growing interest in the local structural units of networks
as well. Small and well defined sub-graphs consisting of a few vertices 
have been introduced as motifs \cite{Alon_1},
whereas somewhat larger units, associated with more highly
interconnected parts are usually called {\em communities},
clusters, cohesive groups, or
modules \cite{Fortunatoreview,Chen_dense,Frago_synt_coord,clustering3}.
These structural sub-units can 
correspond to multi-protein functional units in molecular 
biology \cite{Ravasz02,Spirin_PNAS}, a
set of tightly coupled stocks or industrial sectors in 
economy \cite{Saramaki_stock_Phisica_A}, 
groups of people \cite{Scott_book}, 
cooperative players \cite{Szabo_Phys_Reports}, etc. 
The location of such building blocks can be 
crucial to the understanding of the structural and functional
properties of the systems under investigation. 

The complexity and the size of the investigated data sets are
increasing every year. In parallel, the increasing number of available
computational cores within a single computer or the advent of cloud
computing provides an infrastructure, where such data can be
processed. However, the performance potential of these systems is
accessible only for problems, where the data processing can be
distributed among several computing units. 
Here we introduce the parallel version of CFinder \cite{www},
suitable for finding and visualizing overlapping clusters in large networks. 
This application is based on the earlier, serial 
version of CFinder, which turned out to be a quite popular network
 clustering program. 

The paper is organized as follows. In Section 2 we give a summary of the
Clique Percolation Method (CPM). This is followed by the description of the 
method in Section 3, which distributes the computation among several CPUs or
computing units. The Section 4 is devoted for experimental analysis of the
time complexity of the method. In the last Section we conclude our findings.

\section{The Clique Percolation Method}

Communities are usually defined as dense parts of 
networks and the majority of the community finding approaches 
separate these regions from each other 
by a relatively small number of links in a disjoint manner.
 However, in reality communities may even overlap as well.
In this case the nodes in the 
overlap are members of more than one community. 
A recently introduced, link density-based community
finding technique allowing community overlaps is given by the CPM. 

In this approach a community is built up from adjacent blocks of the same
size $k$. These blocks correspond to $k$-cliques,  corresponding to subgraphs
with the highest possible density: each of the $k$ members of 
the $k$-clique is linked to every other member. Two
blocks are considered adjacent if they overlap with each other as
strongly  as possible, i.e., if they share $k-1$ nodes. Note that
removing one link from a $k$-clique leads to two adjacent 
$k-1$-cliques sharing $k-2$ nodes.

A community is a set of blocks that can be reached from one
to the other through a sequence of adjacent blocks. Note that any
block belongs always to exactly one community, however, there may
be nodes belonging to several communities at the same time. 
A consequence of
 the above definition is that the communities contain only densely
connected nodes. Thus, nodes with only a few connections or not participating
 in a densely connected subgraph are not classified
into any community. We note that the $k$ parameter can be chosen
according to the needs of the user. If one is interested in broader
community covers, then communities at small $k$ values are appropriate.
If the most dense community cores are the target of the study, then the
communities at larger values of $k$ apply. For a general case we recommend
a $k$ value just below the percolation threshold \cite{cpmperc}.
The pseudocode for CPM is given in Algorithm \ref{alg:serCPM}.

\begin{algorithm}
\caption{Community finding by serial CPM}
\label{alg:serCPM}
\begin{algorithmic}[1]
\Require{ Graph}
\Ensure{ $C(k)$: List of communities for all clique sizes $k$ in Graph}
\State {$SC \leftarrow$ empty set for cliques }
\While{ NODE in Graph}
  \State{$TEMPC \leftarrow$ List of maximal cliques containing NODE}
  \State{$SC \leftarrow SC\cup TEMPC$}
\EndWhile
\State{\begin{minipage}[t]{11cm}
       $OVERLAP \leftarrow$ Create overlapping matrix of maximal cliques\\
{\color{white}$OVER$}  $OVERLAP[i,j]$ = set of common nodes of $i$ and $j$, \\
{\color{white}$OVEROVERLAP[i,j]$ =}  where $i,j\in SC$
\end{minipage}
}
\For{$k$=3 {\bf To} size of maximal clique in Graph}
\State{\begin{minipage}[t]{11cm}
$TG(k) \leftarrow $ Create thresholded graph of overlapping maximal cliques: \\
{\color{white}{$TG(k) $}}  
nodes $=\{i\in SC |(\mathrm{size\ of\ }i)>k\}$, \\
{\color{white}{$TG(k) $}} 
edges $= \{(i,j)|i,j\in \mathrm{nodes\ of\ }TG(k),\  OVERLAP[i,j]>=k\}$
\end{minipage}
}
\State{$C(k)\leftarrow$ connected components of $TG(k)$}
\EndFor
\end{algorithmic}
\end{algorithm}

The CPM is robust against removal or insertion of a single
link. Due to the local nature of this approach, such perturbations
can alter only the communities containing at least one of the end
points of a link. (In contrast, for global methods optimizing a 
homogeneously defined quantity, the removal or insertion of
a single link can result in the change of the overall community structure.) 

We note, that beside the mentioned advantages the CPM has certain limits
as well. E.g., if there are not enough cliques in the network
the method will not find any valuable community structure,
whereas for many large overlapping cliques we may easily obtain
a single percolating community for too low $k$ values.
Due to the deterministic nature, the CPM may find communities in a particular
realization of a random network ensemble. In a general case, though,
the members of communities are usually different in each realizations of the 
ensemble, if $k$ is below the percolation threshold.

\begin{figure}[hb]
\centerline{\includegraphics[width=\textwidth]{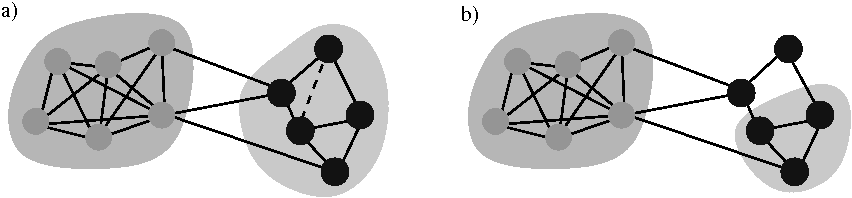}}
\caption{ 
The local nature of the CPM: removing links from the network 
has no effect on the communities which do not contain any of the
endpoints.
The link, which is represented by a dashed line in Fig.\ a) is removed
from the network. The resulting community structure is shown in Fig.\ b). 
The left community (grey nodes) is not affected by the link
removal, since the link is not part of the community. The community on
the right side of the figure (black nodes) is partially effected.
} 
\label{fig:subcl}
\end{figure}

Finally, we point out that the CPM will find the same communities in a
given subgraph irrespective to the fact whether the subgraph is linked
to a larger network or not (see Fig. \ref{fig:subcl}). Therefore, a
heterogeneous network can be 
analyzed by first dividing it into homogeneous parts, and applying the
method to these subnetworks separately.

\section{Distributing the community finding on a grid}

The distributed version of the CPM takes advantage of the local
property of the community definition. Since the communities depend
only on the local network structure, the network can be divided
into small pieces. Then the communities
(or the building blocks for the communities) can be located
in each piece of the network independently.
The distributed CPM is composed of the following main stages:
\begin{enumerate}
\item splitting up the network into pieces,
\item finding communities in each piece of the network,
\item merging the communities found in the previous step.
\end{enumerate}
We provide a pseudocode description in Algorithm \ref{alg:pCPM}.

\begin{algorithm}
\caption{Parallel CPM}
\label{alg:pCPM}
\begin{algorithmic}[1]
\Require{Graph, Size limit for subnetworks $sls$}
\Ensure{List of communities}
\State{$SG\leftarrow$ procedure SPLIT\_OF\_INPUT\_GRAPH(Graph)}
\For{Subgraphs $\in SG$}
\State{$SC\leftarrow$ Collect communities in subgraphs by serial CPM}
\EndFor
\State{\begin{minipage}[t]{11cm}
$OM[i,j]\leftarrow$ Create overlap matrix of communities of subgraphs\\
$i,j\in SC$, $OM[i,j]={\mathrm{size\ of\ largest\ common\ clique\ in\ }i, j}$
\end{minipage}
}
\For{$k\leq\ $ size of largest clique in Graph}
\State{\begin{minipage}[t]{11cm}
$TG(k)\leftarrow$ Create thresholded graph from $OM$\\
where nodes of $TG(k)$ are $k$-clique communities from $SC$\\
and nodes $i,j$ are connected, if $OM[i,j]>=k$
\end{minipage}
}
\State{Communities at $k$ are the union of nodes of connected components of TG(k)}
\EndFor

\Procedure{Split\_of\_input\_Graph}{Graph}
\State{$UN\leftarrow$ nodes in Graph, $LSG \leftarrow$ Initialize list of subgraphs}
\State{Find highest  degree nodes in $UN$\label{state:goto}}
\State{\begin{minipage}[t]{11cm}
$LSG \leftarrow$ add subgraphs containing only a single high degree node
\end{minipage}
}
\While{Size limit is reached for all $sg \in LSG$}
\For{$sg \in LSG$ and size of $sg < sls$}
\State{\begin{minipage}[t]{8cm} Attach nodes to $sg$ 
by one step \\
of a breadth first search process on the input Graph
\end{minipage}
}
\State{$NN(sg) \leftarrow$ nodes attached in this step to $sg$ }
\For{$n\in NN(sg)$}
\If{$\exists o \in LSG$: $o\ne sg$ and $n\in o$ }
\State{\begin{minipage}[t]{8cm}
Flag $n$ in $sg$: do not use this node \\
in further breadth first search steps for subgraph $sg$
\end{minipage}
}
\EndIf
\EndFor
\EndFor
\EndWhile
\For{$sg \in LSG$}
\State{Attach all edges from input Graph among nodes in $sg$ to subgraph $sg$}
\EndFor
\State{$UN\leftarrow$ Nodes in input Graph that are not in any subgraph}
\If{$UN$ is not empty}
\State{Go to line \ref{state:goto}}
\EndIf
\State{Merge small subgraphs in $LSG$ with their neighbors}
\EndProcedure(\Return{LSG})
\end{algorithmic}
\end{algorithm}

\begin{figure}[hb]
\centerline{\includegraphics[width=\textwidth]{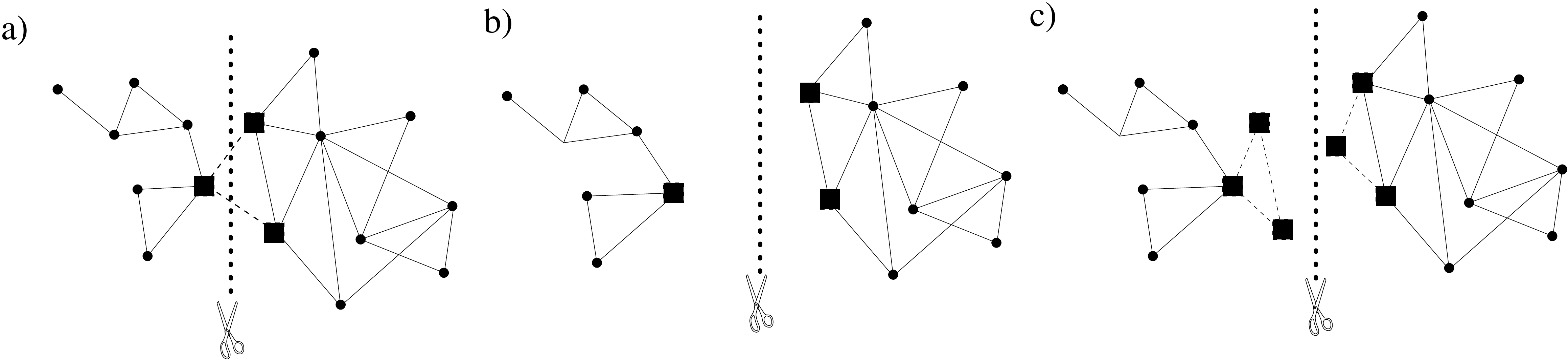}}
\caption{ 
Splitting of a network into two pieces before the exploration of the 
communities. a) The original network: the links highlighted by 
dashed lines are selected as cut-links, their end-nodes define the boundary
 region between the two pieces. b) After the link removal
the network falls into two separate subnetworks. 
c) Re-inserting the boundary region into
 each subnetwork separately. This way the $k$-cliques of the boundary
 region will appear in both pieces.
} 
\label{fig:grsplit}
\end{figure}

The first step is the most crucial one in the process, since 
it has to satisfy the following conditions: 
\begin{itemize}
\item Each part must be sufficiently small to be 
processable by one computing unit.
\item The network should not be split up into too many pieces, since 
the community finding procedure is not optimal on too small networks,
and the computational overhead in the last, merging step becomes too high.
\item Since the splitting step might divide communities as well, 
the nodes at split borders should appear in
 both subnetworks. In the final step these duplicated nodes can be
used to construct the global  community structure from the local 
communities of the subnetworks (see Fig. \ref{fig:grsplit}).
\end{itemize}
The first and the second condition are contradictory: 
if one optimizes for memory usage on a single processing host, the
network has to be split into numerous tiny subnetworks. 
However, as more subnetworks are created, the number of nodes appearing
in mutual split borders is increasing as well,
resulting in inefficient overall memory consumption and CPU usage.
Naturally, the optimal solution depends on the available resources.
As a rule of thumb, one should distribute the
tasks among the processing units such that each unit works with
the largest piece of network processable on the given unit.

The third condition, which requires the ability to reconstruct
the global community structure from the locally found communities 
(and community parts) can be satisfied as follows.
For simplicity let us suppose that we would like to split
 the investigated network into two parts, as shown in Fig. \ref{fig:grsplit}.
First we select a set of links (indicated with dashed lines), whose
removal cuts the network into two separate subnetworks. The end-nodes
of these links (indicated by filled squares) define the boundary region of the
subnetworks. We split 
the network into two pieces by removing the selected links, and for
each subnetwork we separately insert back all nodes and links in the
boundary region (including links between boundary nodes that were not
cut-links) which means that the boundary region is duplicated. As a
result, the $k$-cliques located in the boundary region of the original
network will appear in both subnetworks. Thus, 
 the communities found in the two pieces will overlap in these $k$-cliques,
enabling the reconstruction of the original communities.
The resulting isolated subgraphs can be clustered independently,
therefore, the calculation can be distributed among several
computational units (PCs or processor cores).

Each individual task of the clustering process calculates the
CPM-communities on each subnetwork. Thus for each network piece
the chains of maximally overlapping $k$-cliques are known.
Since a given $k$-clique can be part of only one community, the
communities for the whole network can be built up by merging the
$k$-cliques from the boundary regions of the subnetworks as follows. 

First we build a hyper-network from the network pieces in which 
nodes correspond to subnetworks, and links signal a shared boundary
 region between the subnetworks.  
For each hyper-node we check whether the CPM has found any
communities in the corresponding subnetwork or not.
If communities were found separately for adjacent hyper-nodes,
the overlapping region of the two corresponding subnetworks is checked,
and communities (originally in different subnetworks) sharing a
common $k$-clique are merged.
By iterating over the hyperlinks in this manner, the 
communities of the original network build up from the
merged communities.
Note that the hyperlinks can be processed in parallel,
where the communities are indexed by an array in shared memory or in a
shared database.

\section{Performance analysis and experimental results}

We have tested the method on the two largest example networks available
in the CFinder \cite{www} package: the coauthorship network 
(number of nodes~=~30561, number of links~=~125959) 
and the undirected word association network (number of nodes~=~10627, 
number of links~=~63788).
The main parameter, which has impact on the performance of the algorithm, is 
the size of the
subnetworks the network is split into. Note that this is the minimum
size for a subnetwork. If a large clique is attached to a subnetwork, it cannot
be split up, it is either contained in one subnetwork or it is fully contained
in several subnetworks. 

The parallel version has three main type of computational overhead compared 
to the serial version. The first one is the splitting step, where the
graph is split into smaller pieces. The second source of the processing
overhead is queuing the parallel jobs into a scheduler and 
waiting for free computing 
units. The third one is the merging of the communities from several 
subnetworks. If the total CPU consumption is also an issue, one has to take
into account a fourth type of overhead: 
the processing time of the overlapping network regions, as
these computations are performed more than once. Since the merging step is
implemented as a simple database update command, we measured the time 
consumption of the scheduling and the merging steps together. 

\begin{figure}
\includegraphics[width=63mm]{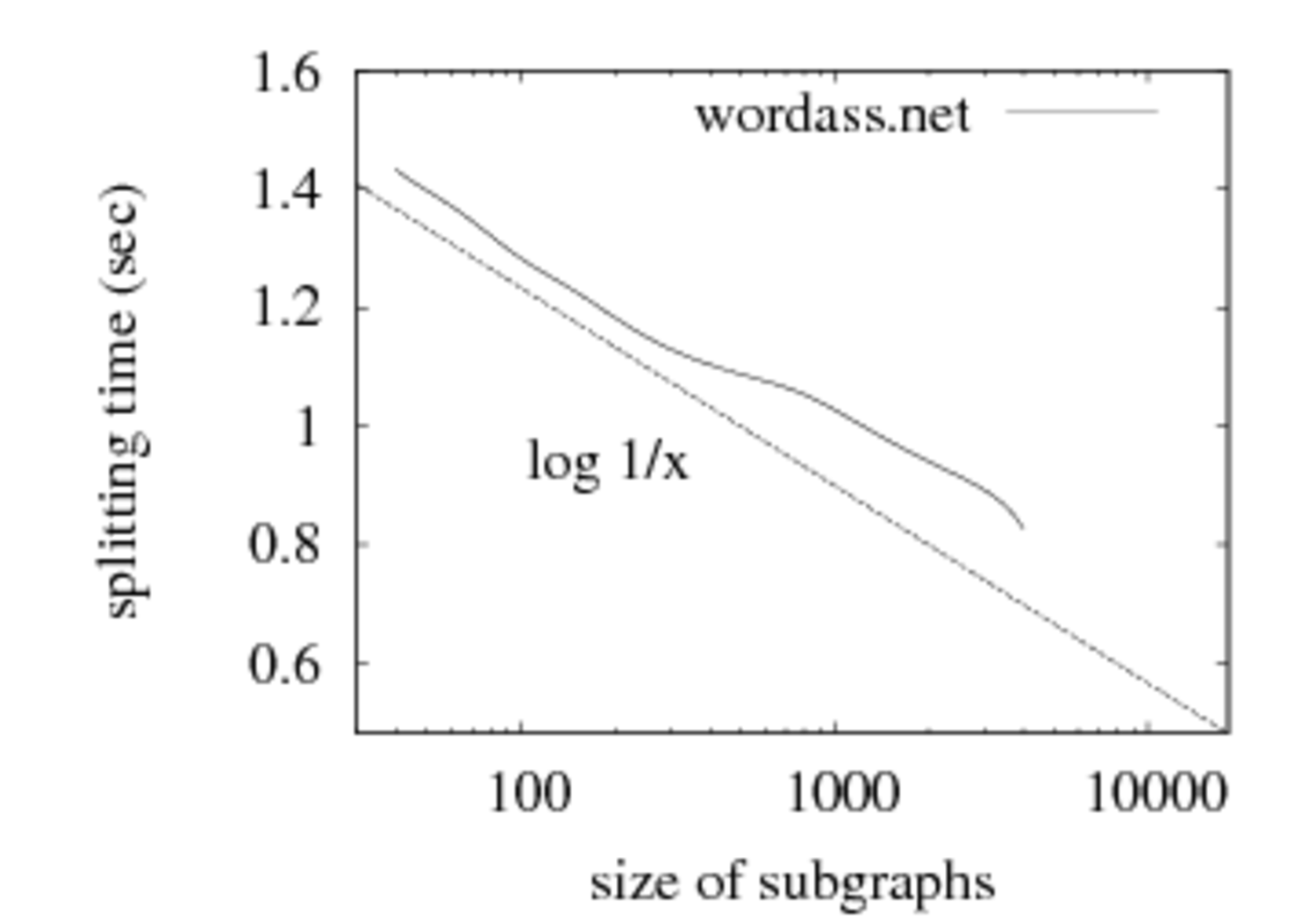}
\includegraphics[width=63mm]{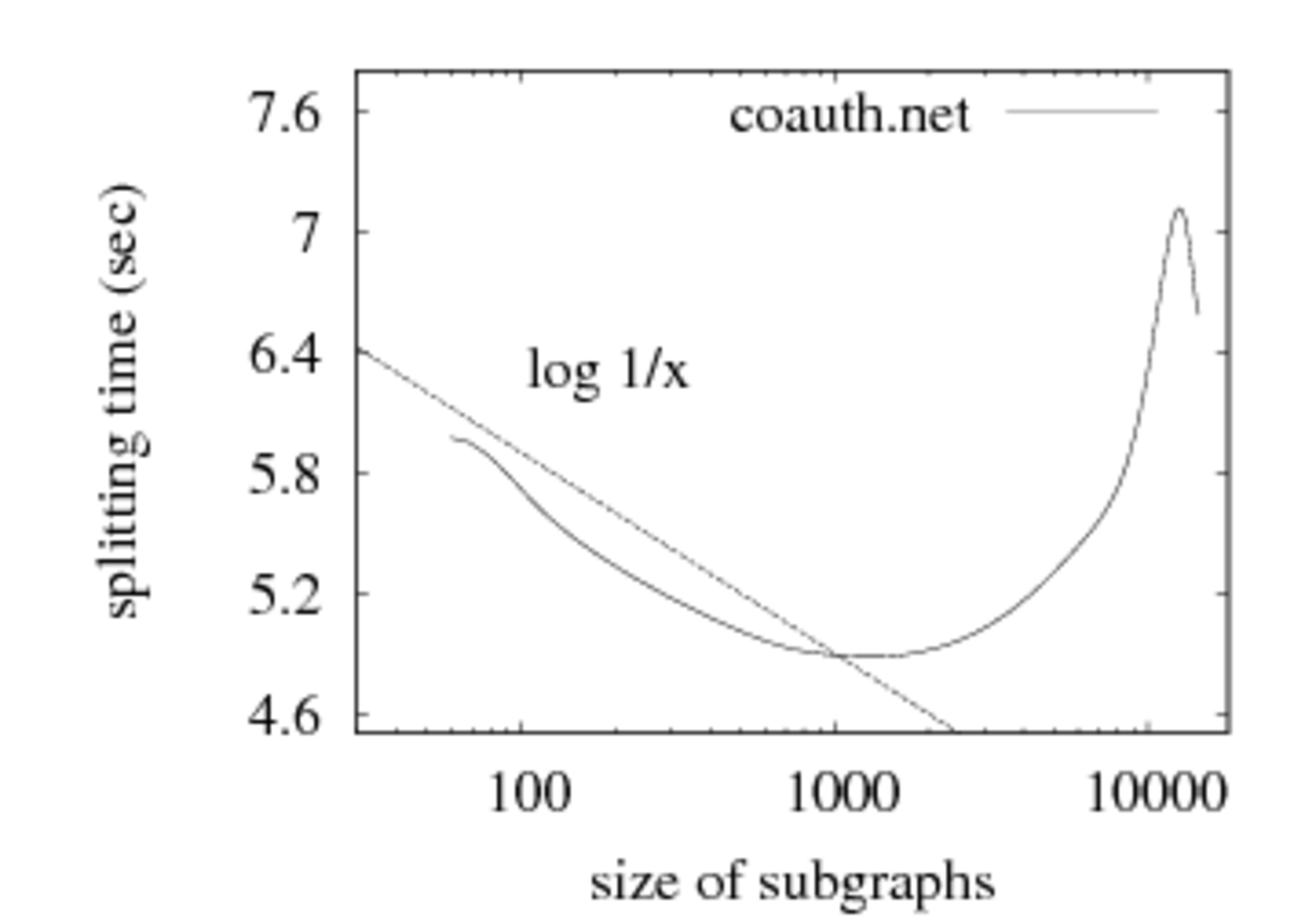}
\caption{Time spent for the splitting step for two example networks. 
Left: word associations network. Right: coauthorship network. The splitting
step is a breadth first search process. The computing time drops 
slowly with increasing subnetwork size. 
A logarithmic decreasing function is
plotted for guiding the eye. The figure on the right demonstrates a case, where
collecting many small subnetworks increases the computing time spent 
for the splitting step.
\label{fig:split}}
\end{figure}

First we analyze the splitting step. In this step the subnetworks are built
up by the breadth first search algorithm. Building a larger network is less 
time consuming than building several small networks, thus, the time complexity
is proportional to the logarithm of the subnetwork size. 
When plotting the splitting time as a function of the subnetwork size 
on a semilogarithmic plot, the slope of its decay is proportional to the 
average branching factor in the breadth first search process
(see Fig. \ref{fig:split}). 

If the high degree nodes
in the network are close to each other, the breadth first algorithm
builds up large subnetworks in a few steps and the size limit for the 
subnetwork is reached. Hence the rest of the nodes outside the large 
subnetworks will
form many disconnected small subnetwork pieces. Our graph splitting algorithm
collects such tiny subnetworks and attaches them to the larger network pieces.
For networks, where this collecting step is needed, the
graph splitting algorithm may consume more computing time 
(see Fig. \ref{fig:split}).

\begin{figure}
\includegraphics[width=8cm]{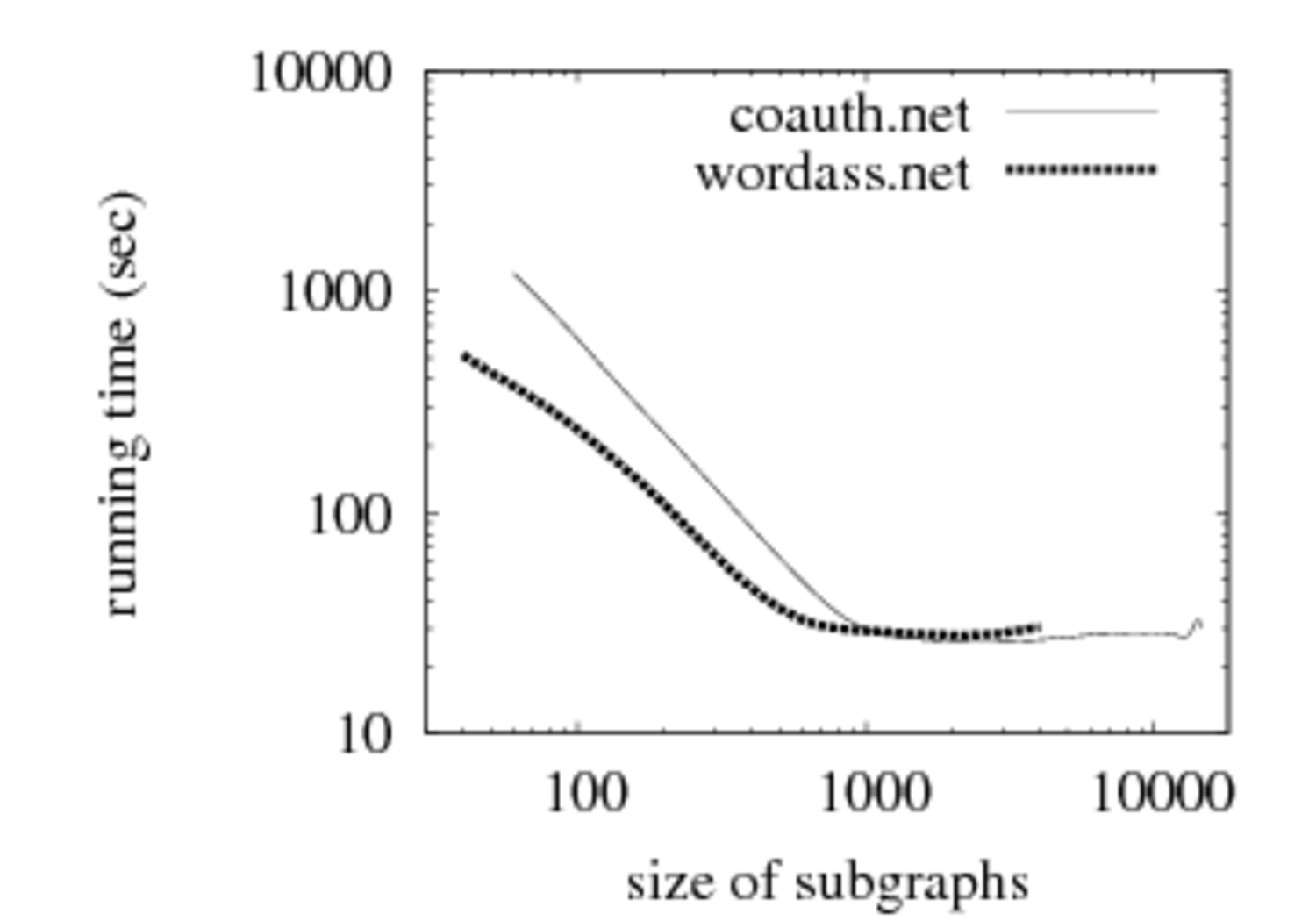}
\caption{The total running time is decreasing with the inverse of the subnetwork
size until the number of subnetworks reaches the number of available processing
units. The optimal size for the subnetworks is the largest possible size, which
fits into one database.
\label{fig:run}}
\end{figure}

Now we turn to the second source of the computing overhead of the 
parallel method. If the network is split up into more subnetworks than
the number of available computing units,
they cannot be processed in parallel, and some jobs must wait until
the previous jobs finish. This effect dominates the running time as shown
on Fig.~\ref{fig:run}. The running time decays linearly with the number of
subnetworks, which is inversely proportional to the subnetwork size. This 
trend is valid until the number of processes reaches the number of available 
computing nodes. Above this subnetwork size the computing time is
practically constant. The faster is the finding of the communities in smaller
subnetworks, the more time is needed for merging the results from various
subnetworks.

In our implementation we used a small grid of personal computers, where
the Condor~\cite{condor} scheduling system distributed the jobs among 
30 cores on linux computers with 2GHz AMD Opteron CPUs connected by
100Mb/s ethernet network.
Here the scheduling time and the communication
overhead among the computing units is comparable to the 
processing time of the largest network, which is manageable in one 
computing unit. In similar environments we advise to use the serial version
for small networks, since the parallel version will not give any
advantage. The main targets of the parallel version are very large networks
that do not fit into the memory of the computers available for the user. We 
note that for typical sparse networks the parallel version will not run faster
on common architectures than the serial version. We expect that for special
networks, where the splitting step results a large number of subnetworks
with negligible number of cliques in the overlapping regions, the parallel 
version can be faster than the serial one provided that
enough computing resources are available, e.g. using GPUs with high 
bandwidth interface.
Such networks are not typical, therefore, our current
implementation is aimed mainly to handle very large networks. 
The size of the processable network is limited be the first 
splitting step and by the last merging step, since here 
the network must be stored either in memory or on disks. If the network
does not fit into the memory it is possible to apply effective disk based
methods in these steps~\cite{parallel_bfs}.

\section{Conclusion}
\label{conclude}

We have presented a parallel implementation of the CFinder \cite{clustering3}
algorithm. We have shown that due to the local nature
of the underlying clique percolation method, the computation
can be distributed among several computational units.
The parallel version may solve large scale network clustering tasks, 
where lacking enough computing resources, e.g. the main memory of the available
computer, would not allow to find the community structure.

\section*{Acknowledgment}
The Project is supported by the European Union and co-financed by the 
European Social Fund (grant agreement no. TAMOP
4.2.1/B-09/1/KMR-2010-0003) and the National Research and Technological
Office (NKTH Textrend).

\end{document}